\documentclass[twocolumn, times, tighten]{aastex62}

\usepackage{graphicx}
\usepackage{epstopdf}
\usepackage[intlimits,sumlimits]{amsmath}
\usepackage{epsfig} 
\usepackage{amssymb}
\usepackage{mathrsfs}  

\usepackage{natbib}

\usepackage{amsmath}
\usepackage{xcolor}
\definecolor{dark-red}{rgb}{0.4,0.15,0.15}
\definecolor{dark-blue}{rgb}{0.15,0.15,0.4}
\definecolor{medium-blue}{rgb}{0,0,0.5}
\hypersetup{
    colorlinks, linkcolor={dark-red},
    citecolor={dark-blue}, urlcolor={medium-blue}
}

\newcommand{\beqa}{\begin{eqnarray}} 
\newcommand{\eeqa}{\end{eqnarray}}

\newcommand{\bsub}{\begin{subequations}}
\newcommand{\esub}{\end{subequations}}
\newcommand{\beal}{\begin{align}}
\newcommand{\ealn}{\end{align}}
\newcommand{\Nif}{$\rm ^{56}Ni$} 
\newcommand{\Cif}{$\rm ^{56}Co$}

\newcommand{\msun}{M$_{\sun}$}
\newcommand{\rsun}{R$_{\sun}$}
\newcommand{\brad}{$R_{\mathrm{BB}}$}
\newcommand{\btemp}{$T_{\mathrm{BB}}$}

\newcommand{\Ni}{\ensuremath{^{56}\mathrm{Ni}}}

\newcommand{\Rsun}{{\ensuremath{\mathrm{R}_{\odot}}}}

\submitjournal{ApJ}

\shortauthors{C.~Fremling et al.}
\shorttitle{ZTF18aalrxas: A Type IIb Supernova from a very extended low-mass progenitor}

\begin{document}

\title{ZTF18aalrxas: A Type IIb Supernova from a very extended low-mass progenitor}

\correspondingauthor{C.~Fremling}
\email{fremling@caltech.edu}

\author{C.~Fremling}
\affiliation{Division of Physics, Mathematics and Astronomy, California Institute of Technology, Pasadena, CA 91125, USA}

\author{H.~Ko}
\affiliation{National Central University, Computer Science and Information Engineering, Taoyuan City 32001, Taiwan (R.O.C.)}

\author{A.~Dugas}
\affiliation{Division of Physics, Mathematics and Astronomy, California Institute of Technology, Pasadena, CA 91125, USA}

\author{M.~Ergon}
\affiliation{Department of Astronomy, The Oskar Klein Center, Stockholm University, AlbaNova, 10691 Stockholm, Sweden}

\author{J.~Sollerman}
\affiliation{Department of Astronomy, The Oskar Klein Center, Stockholm University, AlbaNova, 10691 Stockholm, Sweden}

\author{A.~Bagdasaryan}
\affiliation{Division of Physics, Mathematics and Astronomy, California Institute of Technology, Pasadena, CA 91125, USA}

\author{C.~Barbarino}
\affiliation{Department of Astronomy, The Oskar Klein Center, Stockholm University, AlbaNova, 10691 Stockholm, Sweden}

\author{J.~Belicki}
\affiliation{Caltech Optical Observatories, California Institute of Technology, Pasadena, CA 91125, USA}

\author{E.~Bellm}
\affiliation{ DIRAC Institute, Department of Astronomy, University of Washington, 3910 15th Avenue NE, Seattle, WA 98195, USA}

\author{N.~Blagorodnova}
\altaffiliation{VENI Fellow}
\affiliation{Department of Astrophysics/IMAPP, Radboud University, Houtlaan 4, 6525 XZ  Nijmegen, The Netherlands}

\author{K.~De}
\affiliation{Division of Physics, Mathematics and Astronomy, California Institute of Technology, Pasadena, CA 91125, USA}

\author{R.~Dekany}
\affiliation{Caltech Optical Observatories, California Institute of Technology, Pasadena, CA 91125, USA}

\author{S.~Frederick}
\affiliation{Department of Astronomy, University of Maryland, College Park, MD 20742, USA}

\author{A.~Gal-Yam}
\affiliation{Benoziyo Center for Astrophysics, The Weizmann Institute of Science, Rehovot 76100, Israel}

\author{D.~A.~Goldstein}
\altaffiliation{Hubble Fellow}
\affiliation{Division of Physics, Mathematics and Astronomy, California Institute of Technology, Pasadena, CA 91125, USA}

\author{Z.~Golkhou}
\altaffiliation{Moore-Sloan, WRF Innovation in Data Science, and DIRAC Fellow}
\affiliation{DIRAC Institute, Department of Astronomy, University of Washington, 3910 15th Avenue NE, Seattle, WA 98195, USA}
\affiliation{The eScience Institute, University of Washington, Seattle, WA 98195, USA}

\author{M.~Graham}
\affiliation{Division of Physics, Mathematics and Astronomy, California Institute of Technology, Pasadena, CA 91125, USA}

\author{M.~Kasliwal}
\affiliation{Division of Physics, Mathematics and Astronomy, California Institute of Technology, Pasadena, CA 91125, USA}

\author{M.~Kowalski}
\affiliation{Institut f\"ur Physik, Humboldt-Universit\"at zu Berlin, Newtonstrasse 15, Berlin 12489, Germany}

\author{S.~R.~Kulkarni}
\affiliation{Division of Physics, Mathematics and Astronomy, California Institute of Technology, Pasadena, CA 91125, USA}

\author{T.~Kupfer}
\affiliation{Kavli Institute for Theoretical Physics, University of California, Santa Barbara, CA 93106, USA}

\author{R.~R.~Laher}
\affiliation{IPAC, California Institute of Technology, 1200 E. California Blvd, Pasadena, CA 91125, USA}

\author{F.~J.~Masci}
\affiliation{IPAC, California Institute of Technology, 1200 E. California Blvd, Pasadena, CA 91125, USA}

\author{A.~A.~Miller}
\affiliation{Center for Interdisciplinary Exploration and Research in Astrophysics and Department of Physics and Astronomy, Northwestern University, 2145 Sheridan
Road, Evanston, IL 60208, USA}
\affiliation{The Adler Planetarium, Chicago, IL 60605, USA}

\author{J.~D.~Neill}
\affiliation{Division of Physics, Mathematics and Astronomy, California Institute of Technology, Pasadena, CA 91125, USA}

\author{D.~A.~Perley}
\affiliation{Astrophysics Research Institute, Liverpool John Moores University, Liverpool Science Park, 146 Brownlow Hill, Liverpool L35RF, UK}

\author{U.~D.~Rebbapragada}
\affiliation{Jet Propulsion Laboratory, California Institute of Technology, Pasadena, CA 91109, USA}

\author{R.~Riddle}
\affiliation{Division of Physics, Mathematics and Astronomy, California Institute of Technology, Pasadena, CA 91125, USA}

\author{B.~Rusholme}
\affiliation{IPAC, California Institute of Technology, 1200 E. California Blvd, Pasadena, CA 91125, USA}

\author{S.~Schulze}
\affiliation{Benoziyo Center for Astrophysics, The Weizmann Institute of Science, Rehovot 76100, Israel}

\author{R.~M.~Smith}
\affiliation{Caltech Optical Observatories, California Institute of Technology, Pasadena, CA 91125, USA}

\author{L.~Tartaglia}
\affiliation{Department of Astronomy, The Oskar Klein Center, Stockholm University, AlbaNova, 10691 Stockholm, Sweden}

\author{Lin~Yan}
\affiliation{Caltech Optical Observatories, California Institute of Technology, Pasadena, CA 91125, USA}

\author{Y.~Yao}
\affiliation{Division of Physics, Mathematics and Astronomy, California Institute of Technology, Pasadena, CA 91125, USA}


\begin{abstract}
We investigate ZTF18aalrxas, a double-peaked Type IIb core-collapse supernova (SN) discovered during science validation of the Zwicky Transient Facility (ZTF). ZTF18aalrxas was discovered while the optical emission was still rising towards the initial cooling peak (0.7 mag over 2 days). Our observations consist of multi-band (UV, optical) light-curves, and optical spectra spanning from $\approx0.7$~d to $\approx180$~d past the explosion. We use a Monte-Carlo based non-local thermodynamic equilibrium (NLTE) model, that simultanously reproduces both the \Nif\ powered bolometric light curve and our nebular spectrum. This model is used to constrain the synthesized radioactive nickel mass (0.17~\msun) and the total ejecta mass (1.7~\msun) of the SN. The cooling emission is modeled using semi-analytical extended envelope models to constrain the progenitor radius ($790-1050$~\Rsun) at the time of explosion. Our nebular spectrum shows signs of interaction with a dense circumstellar medium (CSM), and this spetrum is modeled and analysed to constrain the amount of ejected oxygen ($0.3-0.5$~\msun) and the total hydrogen mass ($\approx0.15$~\msun) in the envelope of the progenitor. The oxygen mass of ZTF18aalrxas is consistent with a low ($12-13$~\msun) Zero Age Main Sequence mass progenitor. The light curves and spectra of ZTF18aalrxas are not consistent with massive single star SN Type IIb progenitor models. The presence of an extended hydrogen envelope of low mass, the presence of a dense CSM, the derived ejecta mass, and the late-time oxygen emission can all be explained in a binary model scenario.
\end{abstract}

\section{Introduction}
\label{sec:intro}

The lightcurves (LCs) and spectra of stripped-envelope (SE) supernovae (SNe) can show a wide range of different behaviours. In particular, SNe~IIb show intermittent signatures of hydrogen in their photospheric spectra (e.g., \citealp{1997ARA&amp;A..35..309F,2017hsn..book..195G}). Furthermore, given early enough observations, the presence of hydrogen tends to be accompanied by an initial cooling phase in the optical LCs before the main 
radioactively 
powered peak (e.g., SN~1993J; \citealp{1993Natur.364..600S}; \citealp{1993ApJ...415L.103F}; \citealp{1993Natur.364..507N}, SN~2011dh; \citealp{2011ApJ...742L..18A,2014A&amp;A...562A..17E}; \citealp{2015A&amp;A...580A.142E} and SN~2013df; \citealp{2014AJ....147...37V,2015ApJ...803...40B}).

In some SNe~IIb, the cooling signature dominates the early optical LCs during the first weeks following the explosion (SNe~1993J and 2013df). However, in SN~2011dh, there were strong hydrogen features present in early spectra, but the cooling phase lasted less than 5 days \citep{2011ApJ...742L..18A}. A similar evolution was also seen in e.g., SN~2008ax \citep{2008MNRAS.389..955P,2008MNRAS.391L...5C}, and PTF12os \citep{2016A&amp;A...593A..68F}. The early cooling emission is a result of the SN shock breaking out of the stellar envelope, where the strength
and duration of the optical emission is driven by the radius and mass of the envelope material (e.g., \citealp{2011ApJ...728...63R,2014ApJ...788..193N,2015ApJ...808L..51P}). Thus, an important set of progenitor parameters can be directly probed by studying the early optical emission.

The varying strength of the cooling emission seen in SNe~IIb can be explained if their progenitors have been stripped of their hydrogen envelopes to different degrees. This stripping could either be due to strong stellar winds from very massive stars, with Zero Age Main Sequence (ZAMS) masses, M$_{\mathrm{ZAMS}}>20$~\msun\ (e.g., \citealp{Groh:2013ab}), or due to binary interactions (e.g., \citealp{Yoon:2010aa,2011A&amp;A...528A.131C,2017ApJ...840...10Y}), where the stars do not need to be as massive (typically M$_{\mathrm{ZAMS}}<17$~\msun).

A few SNe~IIb have been found to have broad and slowly evolving lightcurves, indicative of large ejecta masses that could be consistent with massive stars (see e.g., SN~2003bg; \citealp{2009ApJ...703.1612H}, \citealp{2016ApJ...828..111R}, and figure 6 in \citealp{2018A&amp;A...618A..37F}). However, for the majority of SNe~IIb discovered thus far sample studies tend to favor progenitor mass ranges more in line with the expectations for binary models (e.g., \citealp{2011ApJ...741...97D,Cano:2013aa,2015A&amp;A...574A..60T,2016MNRAS.457..328L,2016MNRAS.458.2973P,2018A&amp;A...609A.136T}). Furthermore, for SN~1993J a binary companion has likely been directly observed in post-explosion imaging (\citealp{Maund:2009} and \citealp{2014ApJ...790...17F}), providing the most direct evidence for the binary scenario.

Regardless of the origin of SNe IIb, it is interesting to investigate what the main drivers for the differences in the strengths of the cooling signatures and derived progenitor radii (and extended envelope masses) at the time of explosion are. 
Could it be the case that there is a relation between the ZAMS mass of the progenitor and the mass and extent of the envelopes at the time of explosion? In the case of SN~1993J, nebular models indicate a larger 
ZAMS mass by at least a few solar masses, compared to e.g., SN~2011dh and SN~2008ax \citep{2015A&amp;A...573A..12J}, and SN~1993J has one of the strongest cooling signatures observed in any SN IIb\footnote{SN~2013df also shows a very prominent cooling phase, in addition to a very similar spectral evolution to SN~1993J. \cite{2014MNRAS.445.1647M} argue that a very low ZAMS mass ($12-13$~\msun) could be consistent with the observed oxygen luminosity at late times, but detailed spectral modeling has not been done.}.

In this paper we present a counter example to this ZAMS mass driven picture, and instead argue that the structure of the extended envelope is determined by the binary configuration and subsequent evolution. During science validation the Zwicky Transient Facility (ZTF; \citealp{2019PASP..131a8002B,2019arXiv190201945G}) discovered a SN IIb, ZTF18aalrxas, exceptionally close in time to the explosion. We observed a very strong cooling signature in the optical, almost identical to what was seen in SN~1993J. The cooling phase lasted approximately one week, and was followed by the emergence of a SN IIb lightcurve that qualitatively behaves just like the low mass SN~2011dh. Late time spectra also indicate a similarly low ZAMS mass as seen in SN~2011dh (M$_{\mathrm{ZAMS}}=12-13$~\msun), which puts ZTF18aalrxas among the least massive SNe IIb found to date, while still having a very strong cooling signature.

Our analysis of ZTF18aalrxas spans from $\approx0.7$~days past explosion until $\approx180$~days past explosion. In Sect.~\ref{sec:obs} we describe our follow-up observations and give details on the data reduction involved. Section~\ref{sec:lcs} presents the multi-band LCs and describe the construction of the bolometric LC of ZTF18aalrxas. A qualitative comparison to other SNe IIb shows that the bolometric properties of the SN are not consistent with the progenitor being very massive (i.e., M$_{\mathrm{ZAMS}}>20$~\msun). In Sect.~\ref{sec:spectra} we report our follow-up spectroscopy, confirm the classification of ZTF18aalrxas as a SN IIb and provide velocity measurements of the SN ejecta. In this section we also use late-time spectroscopy ($\approx180$~days past explosion) to constrain the amount of oxygen and hydrogen in the ejecta.
Section~\ref{sec:model} presents a Monte-Carlo based model for the lightcurves and spectra of ZTF18aalrxas, which we use to constrain the synthesized nickel mass and the total ejecta mass of the explosion, along with the helium core mass of the progenitor. Semi analytical models are also used in this section to constrain the mass and radius of the extended envelope. Section~\ref{sec:conclusions} presents our conclusions, and contains a discussion where we put our observations of ZTF18aalrxas in context to other SNe IIb.

\section{Observations}
\label{sec:obs}

\subsection{Detection and classification}
\label{sec:detection}
ZTF18aalrxas was first detected on 2018 April 19.333 ($\mathrm{JD}=2458227.833$), with the Palomar Oschin Schmidt 48-inch (P48) telescope during science validation of the Zwicky Transient Facility \citep{2019PASP..131a8002B,2019arXiv190201945G} and the GROWTH Marshal \citep{2019PASP..131c8003K}. The first detection is in $g$ band, with a host-subtracted magnitude of $19.59\pm0.06$~mag, at the J2000.0 coordinates $\alpha=15^{h}49^{m}11.64^{s}$, $\delta=+32\degr17\arcmin16.8\arcsec$. Observations $\approx22$~h earlier on 2018 April 18 give a pre-explosion limit of $20.69$~mag in $r$ ($5\sigma$ limit computed at the expected position of the transient; $\mathrm{JD}=2458226.899$).

Our first spectrum was obtained on 2018 April 25 with the Palomar 60-inch telescope (P60; \citealp{2006PASP..118.1396C}) with the Spectral Energy Distribution Machine (SEDM; \citealp{2018PASP..130c5003B}). However, this spectrum was largely featureless (with possibly some broad H$\alpha$) and did not lead to a conclusive classification. Follow-up spectra from 2018 April 30 and May 4, showed the emergence of the Balmer series and \ion{He}{1}, characteristic of a SN IIb. Host galaxy emission lines are also present in our spectra at\footnote{The quoted redshift is the average we derive from our 4 Keck I spectra. The error is the standard deviation of these estimates.} $z = 0.0582\pm0.0003$. Thus, we adopt the distance modulus $\mu=37.10$~mag, corresponding to a distance\footnote{Cosmological parameters from WMAP9 \citep{2013ApJS..208...19H}.} of $263$~Mpc to the host galaxy of ZTF18aalrxas.

\subsection{Optical photometry}
\label{sec:optical}
Following the discovery of ZTF18aalrxas, we obtained follow-up photometry during the photospheric phase in $g$ and $r$ with the ZTF Camera on the P48, in $gri$ with the SEDM on the P60, in $ugri$ through the Las Cumbres Observatory Global Telescope Network (LCOGT; \citealp{2013PASP..125.1031B}), in $gr$ with the Kitt Peak Electron Multiplying CCD on the Kitt Peak 84-inch telescope \citep{2019arXiv190104625C}, and in $gri$ with IO:O on the Liverpool Telescope (LT). The Nordic Optical Telescope (NOT) at La Palma was used to obtain late-time photometry in $gri$ around 150 days past the discovery.

Lightcurves from the P48 come from the ZTF production pipeline \citep{2019PASP..131a8003M}, with limiting magnitudes re-computed using forced point-spread function (PSF) fitting at the expected position of the SN (Yao et al., in prep.). Stacked P48 lightcurves were also prepared using \texttt{SkyPortal} \citep{skyportal}, an online interface to the forced PSF-fit photometry. The stacked light curves were produced by taking the inverse-variance weighted average of the single-epoch PSF photometry in non-overlapping 4-day windows. Lightcurves from the rest of our optical imaging data have been produced with the image-subtraction pipeline described in \cite{2016A&amp;A...593A..68F}, with template images from the Sloan Digital Sky Survey (SDSS; \citealp{2014ApJS..211...17A}). This pipeline produces PSF magnitudes, calibrated against SDSS stars in the field. We have also compared our results to PSF magnitudes calibrated against Pan-STARRS1 \citep{2016arXiv161205560C}, finding no significant offsets.

\begin{figure*}
\centering
\includegraphics[width=18cm]{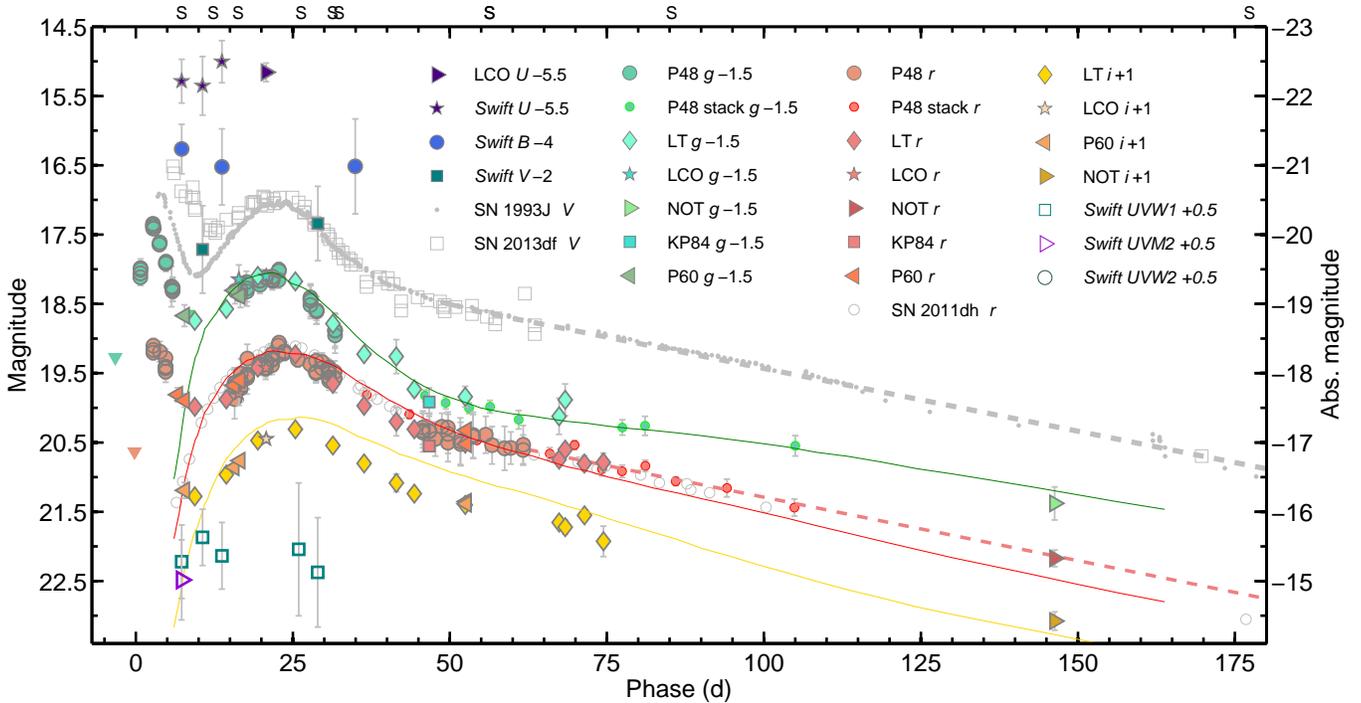}
\caption{Multi-band LCs of ZTF18aalrxas (colored markers). The $V$-band LCs of SN~2013df (gray empty squares) and SN~1993J (gray dots), and the $r$-band LC of SN~2011dh (gray empty circles) are shown as comparisons. The LC of SN~2011dh has been scaled to the distance and nickel mass of ZTF18aalrxas. The LCs of SN~2013df and SN~1993J have been arbitrarily shifted to match our V-band observations of ZTF18aalrxas. The dashed gray line shows a fit to the $V$-band LC of SN~2013df, starting from 50~d past explosion. The dashed red line shows a fit to the $r$-band LC of ZTF18aalrxas, starting from 50~d past explosion. Simulated LCs are shown as solid colored lines (see Sect.~\ref{sec:model}). The stacked P48 LCs contain data from 4 nights of observations for each point. All LCs have been corrected for host-galaxy and MW extinction (see Sect.~\ref{sec:obs}).}
\label{fig:lcs}
\end{figure*}

In our analysis we have corrected all photometry for galactic extinction, using the Milky Way (MW) color excess $E(B-V)_{\mathrm{MW}}=0.0192$~mag toward the position of ZTF18aalrxas \citep{2011ApJ...737..103S}. All reddening corrections are applied using the \cite{1989ApJ...345..245C} extinction law with $R_V=3.1$. No further host galaxy extinction has been applied, since there is no sign of any \ion{Na}{1d} absorption in any of our spectra. The multi-color lightcurves of ZTF18aalrxas are shown in Figure~\ref{fig:lcs}.

\subsection{Swift-UVOT photometry}
\label{sec:uvot}
A set of ultraviolet (UV) and optical photometry ($UVW1$, $UVM2$, $UVW2$ and $UBV$) was obtained with the UV Optical Telescope onboard {\it Swift} ($UVOT$; \citealp{2004ApJ...611.1005G}; \citealp{2005SSRv..120...95R}). Our first {\it Swift-UVOT} observation was performed on 2018 April 25, followed by 7 more observations during the photospheric phase of the supernova, and two late-time observations (2018 Oct 24, 2018 Dec 09), in order to estimate the host galaxy contribution.

Lightcurves from {\it Swift-UVOT} were produced using \texttt{HEAsoft} version 6.25, as described by \cite{2009AJ....137.4517B}, with an aperture radius of 3\arcsec. To estimate the host-galaxy contribution at the location of the SN we take a weighted average of the two latest observations. In general the signal to noise ratio (SNR) after host subtraction was very low for this dataset. Only datapoints with SNR $>2$ after host subtraction are considered in our analysis.

\subsection{Optical spectroscopy}
\label{sec:opticalspectra}
Spectroscopic follow-up of ZTF18aalrxas started on 2018 April 25 ($\approx6$~d past discovery) with a spectrum from SEDM mounted
on the P60. Further spectra were obtained with the NOT using the Andalucia Faint Object Spectrograph (ALFOSC),with the Keck-I telescope using the Low Resolution Imaging Spectrograph (LRIS; \citealp{1994SPIE.2198..178O}), with the Dual Imaging Spectrograph (DIS) on the Apache Point Observatory 3.5-meter telescope, and with the Device Optimized for the LOw RESolution (DOLORES) on Telescopio Nazionale Galileo (TNG). Our latest spectrum was obtained on 2018 Oct 12 ($\approx180$~d past discovery) with LRIS on Keck-I.

The \texttt{LPipe} reduction pipeline\footnote{\href{http://www.astro.caltech.edu/dperley/programs/lpipe.html}{http://www.astro.caltech.edu/dperley/programs/lpipe.html}} (Perley et al. in prep) was used to process the LRIS data. The other spectra were reduced using standard pipelines and procedures for each telescope and instrument. All spectral data and corresponding information will be made available via WISeREP\footnote{\href{https://wiserep.weizmann.ac.il/}{https://wiserep.weizmann.ac.il/}} \citep{Yaron:2012aa}.

\section{Light Curves}\label{sec:lcs}
The P48 $g$- and $r$-band pre-explosion limits along with the UV and optical LCs of ZTF18aalrxas are displayed in Fig.~\ref{fig:lcs}.
Qualitatively the behavior of ZTF18aalrxas is well matched to both SN~1993J and SN~2013df during both the first and second LC peaks. The second peak is also well matched to SN~2011dh, even though SN~2011dh lacked a strong initial cooling phase. For the explosion date of ZTF18aalrxas we have adopted $t_{\mathrm{exp}}=2458227.1\pm0.1$~(JD), from our best-fitting model for the early cooling emission\footnote{This value is in good agreement with the result of fitting the early $g$-band flux with a power-law.} (Sect.~\ref{sec:early_lcs}). The decline rates of the LCs after 50~d past explosion also appear very similar for these objects (dashed lines in Fig.~\ref{fig:lcs}). We measure the $r$-band decline rate of ZTF18aalrxas to $1.83\frac{\mathrm{mag}}{100~\mathrm{d}}$ by fitting a first-order polynomial to the data starting from 50~d past explosion.

A pseudo-bolometric LC for ZTF18aalrxas was constructed by fitting a black-body (BB) model to the spectral energy distribution (SED) derived from the lightcurves as a function of time, and integrating the flux of the fitted BB (starting from 2600~\AA). We have weighted our BB fits by the photometric errors. This de-emphasizes the data collected by {\it Swift}, which has very large uncertainties. The photometry has been interpolated to the dates of the $r$-band data points, in order to allow the construction of the bolometric LC. Errors have been calculated using a Monte-Carlo (MC) method where the BB fits are re-computed for the range of lightcurves that are allowed by shifting the LC points within the photometric uncertainties (we assume a normal distribution when resampling).

\begin{figure}
\centering
\includegraphics[width=\columnwidth]{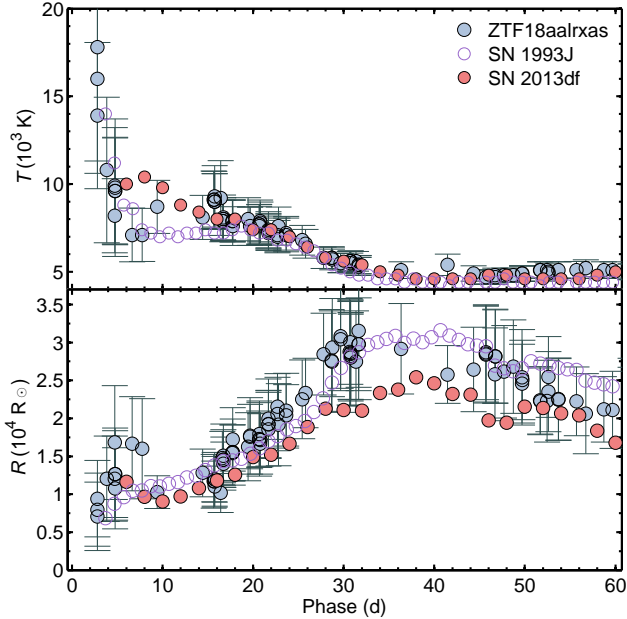}
\caption{Black-body temperature (top panel) and black-body radius (bottom panel) derived from BB fits to the photometry of ZTF18aalrxas (blue circles), SN~1993J (purple unfilled circles), and SN 2013df (red circles).}
\label{fig:btempbrad}
\end{figure}

\begin{figure*}
\centering
\includegraphics[width=14cm]{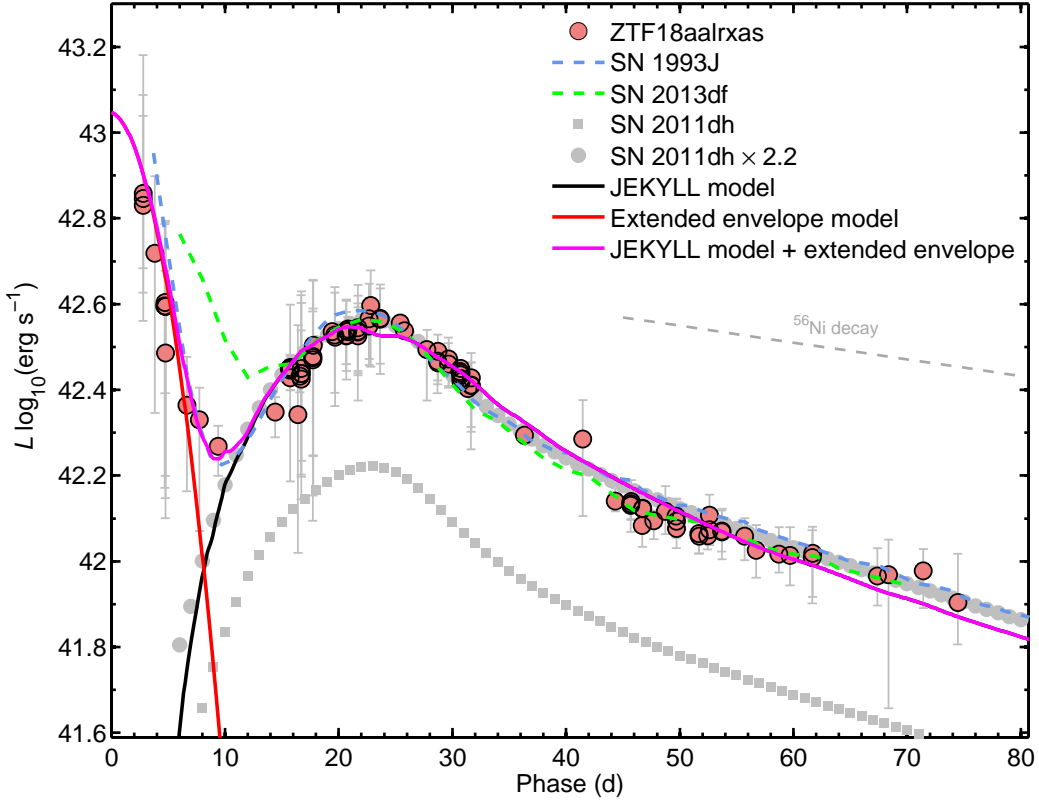}
\caption{Bolometric LC of ZTF18aalrxas (red circles), compared to the bolometric LCs of SN~2011dh (gray circles and squares), SN~2013df (dashed green line) and SN~1993J (dashed blue line). The LCs of SN~2011dh, SN~2013df, and SN~1993J have been scaled to match the nickel mass of ZTF18aalrxas. The bolometric LC from our NLTE simulation is shown as a black line. A best-fit semi-analytic model for the cooling emission from an extended envelope \citep{2016A&amp;A...589A..53N} is shown as a red line (see Sect.~\ref{sec:model}). The sum of the NLTE model and the extended envelope model is shown as a pink line.}
\label{fig:bol_lc}
\end{figure*}

The BB parameters and the pseudo-bolometric LC are shown in Fig.~\ref{fig:btempbrad} and Fig.~\ref{fig:bol_lc}, respectively. In Fig.~\ref{fig:bol_lc} we also display the pseudo-bolometric LC of SN~2011dh (data from \citealp{2014A&amp;A...562A..17E}) both without scaling and with scaling to match the peak luminosity (nickel mass) of ZTF18aalrxas, as well as scaled pseudo-bolometric LCs of SN~2013df (data from \citealp{2016MNRAS.460.1500S}) and SN~1993J (data from \citealp{1994AJ....107.1022R}). The bolometric LCs of SN~2011dh and SN~2013df have been constructed in the same way as the bolometric LC of ZTF18aalrxas. When determining the BB parameters and bolometric LC of SN~2013df we have applied a color-based extinction correction with $E(B-V) = 0.175$~mag, based on the observed $g-r$ color 10~d past the $r$-band peak, as described in \cite{2015A&amp;A...574A..60T}. Since the explosion time for SN~2013df is uncertain, we have matched the phase of maximum light of the main peak in the bolometric LC to that of ZTF18aalrxas\footnote{The light curve comparison based explosion time estimate for SN~2013df by \cite{2014AJ....147...37V} would shift the bolometric LC such that the main peak happens a few days later compared to ZTF18aalrxas.}. For SN~1993J there are no $gri$ data. Instead we have detrmined the BB parameters and constructed the pseudo-bolometric LC via BB fitting of $BVRI$ LCs. Since SN~2013df has both $gri$ and $BVRI$ coverage, we have checked that the bolometric LC does not significantly change depending on the filter set in our method. We have adopted the lower value for the extinction ($E(B-V)=0.08$~mag) used by \cite{1994AJ....107.1022R} when performing BB fits to the SED of SN~1993J.

The peak luminosity of ZTF18aalrxas is higher by about a factor of 2.2 compared to SN 2011dh, indicating a \Nif\ mass\footnote{Some \Ni-mass estimates for SN~2011dh are $0.075\pm0.025$~\msun\ by \cite{2014A&amp;A...562A..17E} and 0.07~\msun\ by \cite{2013MNRAS.436.3614S}.} that should be higher by a similar amount, which turns out to be $\approx0.17$~\msun. This is very close to the median nickel mass (M$_{^{56}\mathrm{Ni}}=0.15\pm0.07$~\msun) of the SN IIb sample studied by \cite{2018A&amp;A...609A.136T}.
The initial cooling phase of ZTF18aalrxas remains very prominent in the bolometric LC, and there is a striking similarity to the cooling phase seen in SN~1993J. The luminosity at the initial peak is also similar to that observed in SN~2013df, but the duration of the cooling phase is shorter in ZTF18aalrxas. After about one week past explosion, the main \Nif-powered LC peak starts to dominate in ZTF18aalrxas, while the cooling dominates for almost one more week in SN~2013df. This indicates a higher mass in the envelope of the progenitor to SN~2013df compared to the progenitor of ZTF18aalrxas (see Sect.~\ref{sec:early_lcs}).

The width of the main bolometric LC peak appears to be very similar for ZTF18aalrxas, SN~1993J, SN~2013df, and SN~2011dh. A bolometric LC well matched to SN~2011dh strongly disfavors a progenitor scenario for ZTF18aalrxas where the mass loss was dominated by the stellar wind from a very massive star, since the LCs of SN~2011dh are consistent with stars with M$_{\mathrm{ZAMS}}\leq13$~\msun. Under the assumption that a compact neutron-star remnant is formed, a massive progenitor (M$_{\mathrm{ZAMS}}\gtrsim20$~\msun) should have a large ejecta mass and slower evolution of the bolometric LC.

From the BB fits to the SED of ZTF18aalrxas we have derived the evolution of \brad,
which can be interpreted as a rough approximation of the photospheric
radius (bottom panel of Fig.~\ref{fig:btempbrad}). \brad\ evolves in a very similar way when compared to SN~2013df, and especially when compared to SN~1993J (shown as comparisons). Our earliest measurement is at 2.8~d past explosion, and indicates \brad\ in the range $0.8\pm0.5\times10^4$ \rsun, which later peaks around $2.5-3\times10^4$ \rsun\ at $\approx40$~d past explosion. \btemp\ (top panel of Fig.~\ref{fig:btempbrad}) starts out at around $15\times10^3$~K in ZTF18aalrxas at 2.8~d past explosion, but drops quickly to around $7.5\times10^3$~K over the following few days. Again remarkably similar to the evolution of SN~1993J, including the rapid decline seen initially. In SN~2013df \btemp\ stayed around $10^4$~K for the first 10~d past explosion. Compared to the larger
sample of SE~SNe presented by \cite{2018A&amp;A...609A.136T}, both
\btemp\ and \brad\ of ZTF18aalrxas appear to be very consistent with the averages for SE~SNe.

\section{Spectra}
\label{sec:spectra}

\begin{figure*}
\centering
\includegraphics[width=18cm]{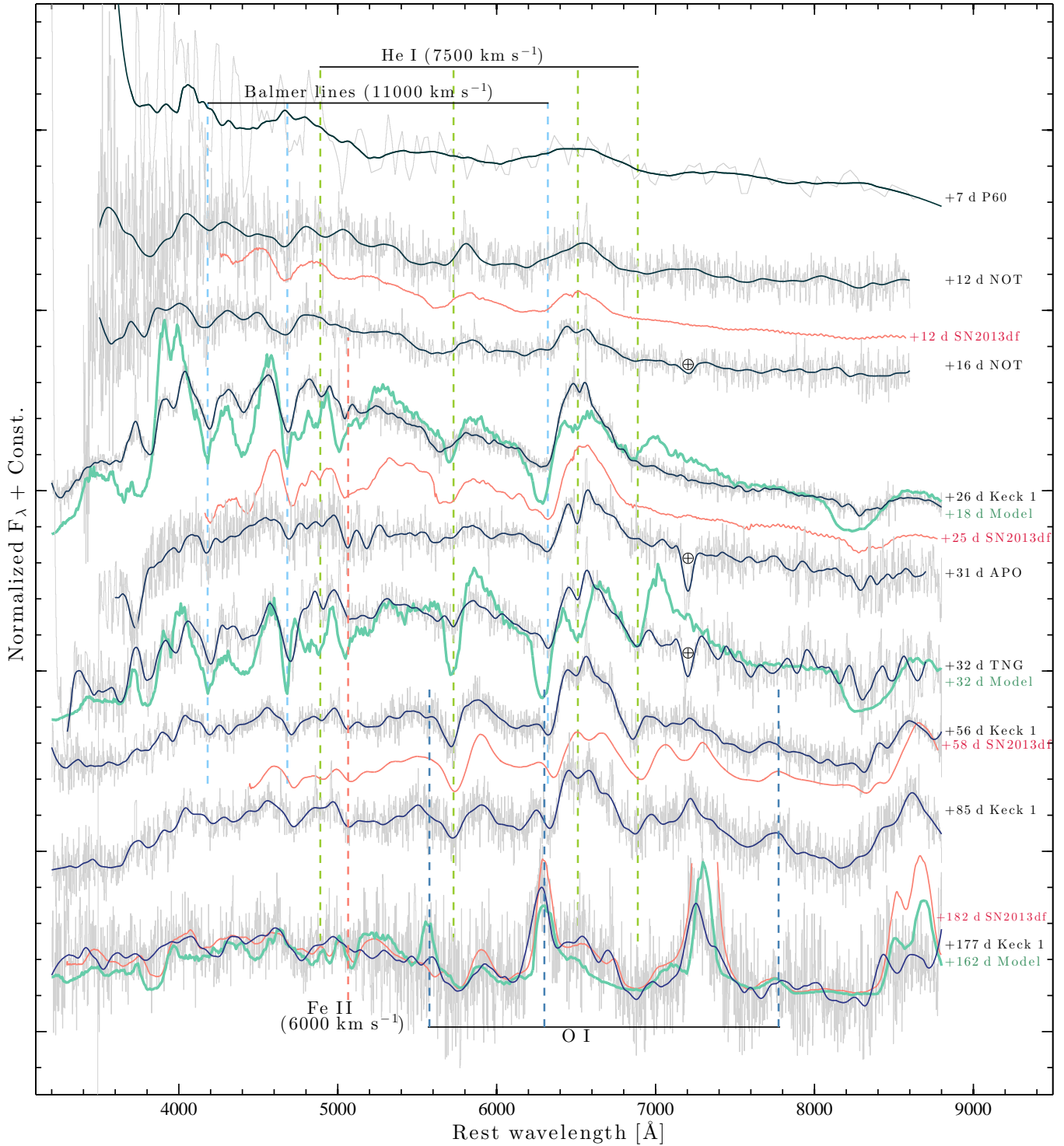}  
\caption{\label{fig:spec_evol}Spectral evolution of ZTF18aalrxas (black) in the wavelength range $3200-8800$~\AA. Comparison spectra of SN~2013df (from \citealp{2016MNRAS.460.1500S,2019MNRAS.482.1545S}) are shown as red lines. Green lines show simulated spectra (see Sect.~\ref{sec:spectra}). Dashed lines mark the central wavelength of the marked emission lines at rest, or shifted blueward to the velocity specified within parentheses to match the absorption minima associated with the emission lines. The \ion{Ca}{2} emission around $7300$~\AA\, has been clipped in the spectrum of SN~2013df. A telluric feature near 7200~\AA\ is present in the spectra taken at $+16$~d, $+31$~d, and $+32$~d.}
\end{figure*}

Our spectral sequence (Fig.~\ref{fig:spec_evol}) starts out at $7$~d past explosion with spectra dominated by the Balmer series. Our earliest spectrum from SEDM is of low SNR, but does show a broad feature where H$\alpha$ is expected, which is also consistent with the NOT spectra taken at $12$~d and $16$~d past explosion, where several of the Balmer lines are present. A spectrum of SN~2013df taken at $12$~d past explosion is qualitatively very similar, but slightly bluer, consistent with the higher \btemp\ we derived in Sect.~\ref{sec:lcs}.

\ion{He}{1} signatures appear rather late in ZTF18aalrxas; we observe the simultaneous appearance of \ion{He}{1} $\lambda\lambda$5016, 5876, 6678, 7065 in absorption at around 7\,500~km~s$^{-1}$ in our spectra past 30~d, with the signatures being quite clear at $56$~d past explosion. This evolution is similar to that of SN~2013df and SN~1993J, which also lacked clear \ion{He}{1} for the first month. In particular, our LRIS spectrum of ZTF18aalrxas taken at $56$~d past explosion is qualitatively very similar to a spectrum taken at a similar epoch of SN~2013df. This clearly solidifies the SN IIb classification of ZTF18aalrxas.

Our spectrum taken at 85~d past explosion shows the emergence of  [\ion{O}{1}] $\lambda\lambda$5577, 6300, 6364 emission lines along with the \ion{O}{1} $\lambda$7774 triplet, hinting that the ejecta are becoming optically thin. The latest spectrum, taken at $177$~d past explosion, is clearly in the nebular phase, with prominent [\ion{O}{1}] $\lambda\lambda$6300, 6364 seen in emission. There is also broad flat-topped H$\alpha$ emission emerging, similar to what was seen in SN~2013df and SN~1993J. This nebular spectrum is further analyzed and modeled in Sect.~\ref{sec:oi} in order to constrain the O and H envelope masses.

\ion{Fe}{2}~$\lambda$5169 is a decent tracer for the photospheric velocity \citep{Dessart:aa}. Absorption from this line is present in our spectra taken later than 26~d past explosion. Velocities measured for \ion{Fe}{2}~$\lambda$5169, H$\alpha$, H$\beta$ and \ion{He}{1}~$\lambda\lambda$5876, 7065 are shown in Fig.~\ref{fig:line_vel}. These measurements were performed by smoothing the spectra and locating the relevant absorption minima. Uncertainties were estimated through a MC simulation where many simulated spectra were created, using smoothed spectra and randomly generated noise reflecting the observed SNR. The \ion{Fe}{2}~$\lambda$5169 velocity of ZTF18aalrxas is in good agreement with the velocity derived from the evolution of the BB radius (Sect.~\ref{sec:lcs}), which gives $v_{\mathrm{ph}}\approx8\,000$~km~s$^{-1}$ around 26~d past explosion. This velocity is also similar to what can be seen in SN~1993J, SN~2013df, and SN~2011dh at a similar epoch.

\subsection{Nebular spectroscopy and oxygen mass constraints}
\label{sec:oi}
More massive stars produce larger amounts of metals, and the oxygen nucleosynthesis in particular is a strong and monotonic function of $M_{\mathrm{ZAMS}}$ \citep{2007PhR...442..269W}. This has been exploited by \cite{2015A&amp;A...573A..12J}, who used a grid of simulated spectra to show a strong dependence between the ratio of the [\ion{O}{1}]~$\lambda\lambda$6300, 6364 line luminosity compared to the total $^{56}$Co decay power, for SE SNe in the nebular phase.

We have used the model grid of \cite{2015A&amp;A...573A..12J} to match our nebular spectrum of ZTF18aalrxas taken at $177$~d past explosion, with a focus on the [\ion{O}{1}]~$\lambda\lambda$6300, 6364 emission (Fig.~\ref{fig:nebular}). The same model that was used by \cite{2015A&amp;A...573A..12J} to model the nebular emission of SN~2011dh (model 12C in \citealp{2015A&amp;A...573A..12J}) is also quite well matched to ZTF18aalrxas. The peak luminosity of the [\ion{O}{1}]~$\lambda\lambda$6300, 6364 line is slightly higher in our observation, but not as high as what would be expected for a $M_{\mathrm{ZAMS}}=13$~\msun\ model. Thus, we conclude that the O mass should be in the range $0.3-0.5$~\msun\ (O mass values from \citealp{2015A&amp;A...573A..12J}), which for these models indicate $M_{\mathrm{ZAMS}}=12-13$~\msun\ for the progenitor of ZTF18aalrxas.

\subsection{Hydrogen mass constraints}
\label{sec:ha}
Our spectrum of ZTF18aalrxas taken at 177~d past explosion shows a clear broad excess emission, around the region of H$\alpha$, when compared to our model spectra (Fig.~\ref{fig:nebular}). Furthermore, the profile of this emission is flat, almost identical to what was seen in SN~1993J and SN SN~2013df at a similar epoch (see Fig.~\ref{fig:spec_evol}). This feature gradually becomes clearer in later spectra of SN~1993J and SN~2013df, and eventually completely dominates the H$\alpha$ region after around 600~d past explosion (e.g., \citealp{2015ApJ...807...35M}). It is not possible to explain the luminosity and broadness of this line at such late epochs with a \Cif\ powered nebular model such as \citealp{2015A&amp;A...573A..12J}. The flat and broad profile indicates that the emitting region has a shell-like geometry, and that interaction with circumstellar material (CSM) is the powering mechanism \citep{1994ApJ...420..268C}.

If we combine the 12C model spectrum from \cite{2015A&amp;A...573A..12J}, with a spectrum of SN~1993J taken at 973~d past explosion (where there is very little contribution from \Cif\ powered metal lines remaining), we find that the excess broad emission around H$\alpha$ can be almost perfectly reproduced. Furthermore, this combined (12C$+$SN~1993J) model is at the same time generally better matched to the observed spectrum in the entire blue part of the spectrum ($\lambda<6000$~\AA). By integrating the broad H$\alpha$ in the spectrum of SN~1993J in the combined 12C$+$SN~1993J model after it was matched to ZTF18aalrxas at 177~d past explosion we estimate the H$\alpha$ luminosity of ZTF18aalrxas to $L_{H\alpha}\approx1.22\times10^{39}$~erg~s$^{-1}$ 
at this epoch\footnote{The accuracy of this estimate is motivated by the fact that there is little velocity evolution in the red and blue edge velocities of the broad H$\alpha$ feature between 171~d to 1766~d past the explosion in SN~1993J \citep{1995A&amp;A...299..715P,2000AJ....120.1487M,2000AJ....120.1499M}.}.

\cite{1995A&amp;A...299..715P} have constructed a model to estimate the H mass of SN~1993J based on the H$\alpha$ luminosity (see their Eq. 2). This equation applied to their H$\alpha$ luminosities between 171~d to 367~d past explosion result in virtually the same upper limit for the H envelope mass for all epochs ($M_H\approx0.2$~\msun). Since the mass estimate is not increasing over time, this favors a scenario where the emitting region is the unshocked ejecta in the CSM model; the entire hydrogen envelope in the ejecta is contributing to the H$\alpha$ luminosity (see also \citealp{2005ApJ...622..991F}). A detailed discussion on SN~2013df along similar lines can be found in \cite{2015ApJ...807...35M}, who were also able to constrain the composition of the emitting region to $40\%$ H and $60\%$ He in mass (which gives the electron density $1.4n_\mathrm{H^{+}}$). If this electron density is applied to SN~1993J, the envelope mass estimate becomes $M_H\approx0.14$~\msun. For SN~2013df \cite{2015ApJ...807...35M} derive $M_H\approx0.2$~\msun. If we apply Eq. (2) in \cite{1995A&amp;A...299..715P} to ZTF18aalrxas, assuming the electron density of SN~2013df, and the H$\alpha$ velocity at the red edge of the line ($v_e=11,500$~km~s$^{-1}$, measured from our nebular spectrum), we find $M_H\approx0.15$~\msun, which gives the total envelope mass (H+He) $M_{env}\approx0.38$~\msun.

In conclusion, the hydrogen envelope masses of SN~1993J and ZTF18aalrxas appear to be very similar, while the hydrogen envelope mass of SN~2013df is somewhat larger. This finding is consistent with SN~2013df having a longer lasting early cooling phase (see Sect.~\ref{sec:early_lcs}). We also note that since the CSM contribution to the H$\alpha$ line appears to be significant already at $\sim170$~d past explosion in these SNe\footnote{\cite{2018ApJ...864...47F} have argued that at $\sim200$~d past explosion, the H$\alpha$-like structure in most SNe IIb is predominately powered by radioactive decay. However, we have here removed the contribution from a radioactively powered SN using the C12 model from \cite{2015A&amp;A...573A..12J}; we consider the excess that cannot be explained by the model as CSM powered emission.}, a dense CSM is required, and they must be exploding when strong mass loss is ongoing; for SN~2013df, \cite{2015ApJ...807...35M} estimate that during the final $\sim800$ years before the explosion the mass-loss rate was $\sim5\times10^{-5}$~\msun~yr$^{-1}$. For SN~1993J a mass-loss rate of $\sim4\times10^{-5}$~\msun~yr$^{-1}$, was derived by \cite{1996ApJ...461..993F}, based on X-ray and radio observations. Given the similarity of ZTF18aalrxas to these objects in terms of the late-time H$\alpha$ line produced by CSM interaction, a comparable mass loss rate would be expected.

\begin{figure}
\centering
\includegraphics[width=8.9cm]{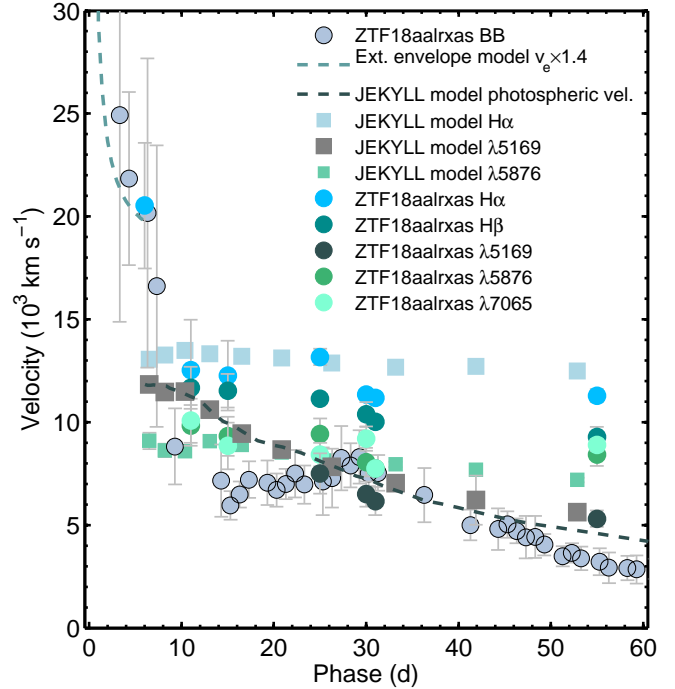}
\caption{Line velocities derived from absorption minima in the spectra of ZTF18aalrxas (colored circles) and our simulated spectra from \texttt{JEKYLL} (colored squares). The expansion velocity derived from BB fits to the photometry of ZTF18aalrxas is shown as steel blue circles with a black outline. The photospheric velocity of our extended envelope model (based on \citealp{2014ApJ...788..193N}) is shown as a dashed blue line, and the photospheric velocity in our \texttt{JEKYLL} model is shown as a dashed gray line.}
\label{fig:line_vel}
\end{figure}

\begin{figure*}
\centering
\includegraphics[width=16cm]{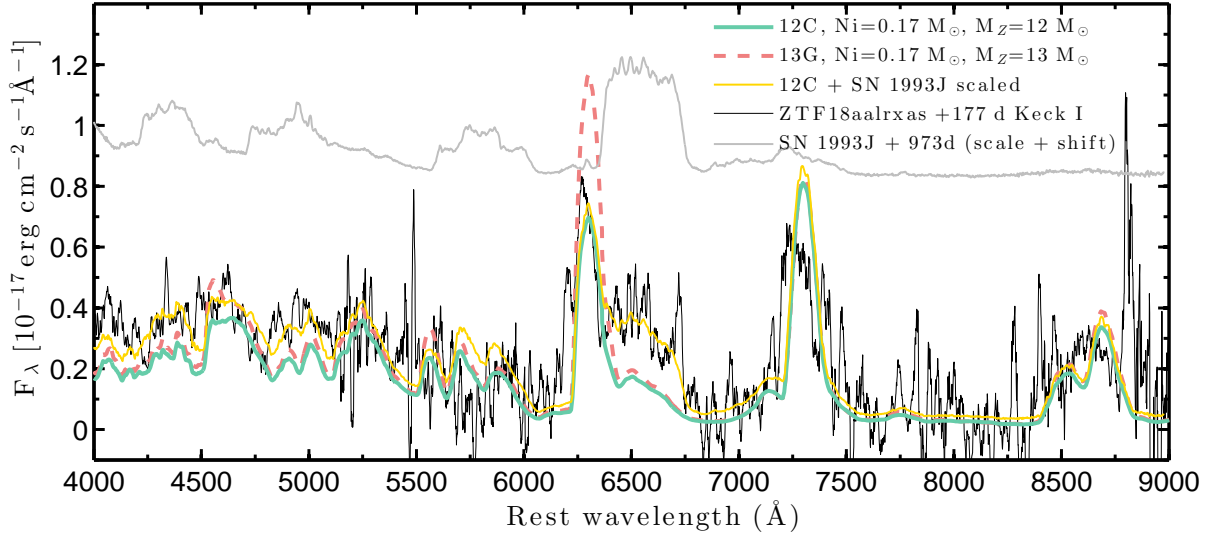}
\caption{Nebular spectrum of ZTF18aalrxas obtained at +177~d (black line) compared to simulated spectra. One simulation is shown for $M_{\rm ZAMS}=13$~\msun\ (red dashed line) and another for $M_{\rm ZAMS}=12$~\msun\ (green line). The observed spectrum is consistent with $M_{\rm ZAMS}$ in the range $12-13$~\msun. The spectrum of ZTF18aalrxas has been flux calibrated using our $r$-band photometry (extrapolated to $177$~d past explosion). All simulations have been scaled to the epochs of the spectra according to the prescriptions of \cite{2015A&amp;A...573A..12J}.}
\label{fig:nebular}
\end{figure*}

\section{Modeling and Progenitor Constraints}
\label{sec:model}

\subsection{Modeling the cooling emission}\label{sec:early_lcs}
\begin{figure}
\centering
\includegraphics[width=8cm]{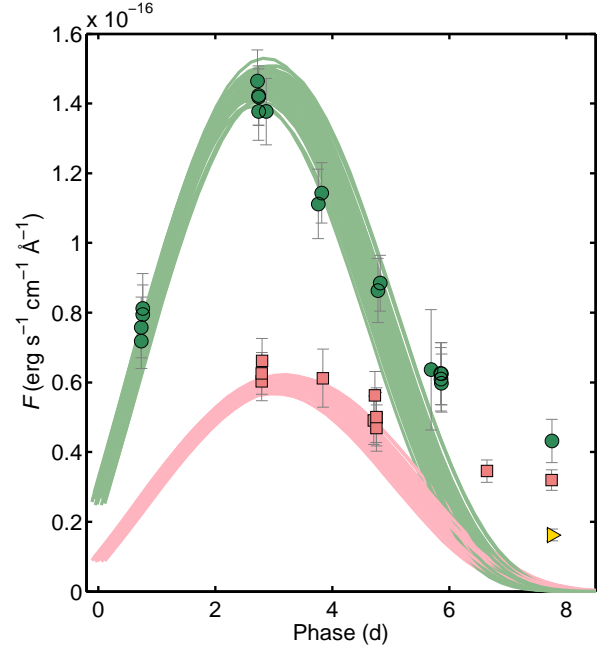}
\caption{Early $gri$ LCs of ZTF18aalrxas (colored markers). The range of models from our MCMC simulation (solid lines) based on the model by \cite{2014ApJ...788..193N}. The range of the models spans the 68\% confidence intervals in the MCMC simulation.}
\label{fig:early_lcs}
\end{figure}

Figure.~\ref{fig:early_lcs} shows the $gri$-band LCs of ZTF18aalrxas up until $\approx8$~d past explosion. We detected ZTF18aalrxas while the $g$-band luminosity was still rising towards the initial peak in the LC. This early emission can be modeled as the result of the SN shock breaking out from a progenitor surrounded by an extended envelope (e.g., \citealp{2011ApJ...728...63R,2014ApJ...788..193N,2015ApJ...808L..51P}). In these models, the SN shock heats the extended material to very high temperatures ($T>10^5$~K), after which the material rapidly starts to cool. This means that what looks like an initial rise to a peak in the optical is actually a temperature effect. The SED of the SN gradually moves into the optical range as the ejecta cool, resulting in rising LCs, but the total bolometric luminosity is actually declining. Regardless, having datapoints on the optical rise to the first peak is a highly significant observation, since it allows a very accurate explosion time estimate, which is rare for double-peaked SNe IIb. For SN~2016gkg, which had a comparable cooling phase duration to SN~2011dh, there is an extremely early discovery ($0.1-0.2$~d past the explosion; \citealp{2018Natur.554..497B}). However, in both SN~1993J and SN~2013df the first points on the LCs are on the decline from the initial LC peaks.

In order to estimate the properties of the extended material that gives rise to the initial LC peak of ZTF18aalrxas we have used the Markov Chain Monte-Carlo (MCMC) method in the Python \texttt{emcee} package \citep{2013PASP..125..306F}, and the model by \cite{2014ApJ...788..193N}. This is a one-zone model for the extended envelope, with the mass in the envelope $M_{env}$ concentrated at a radius $R_{env}$. For the opacity of the envelope material we assume $\kappa=0.3$~cm$^2$g$^{-1}$. The core mass, $M_c$, the time of explosion $t_{\mathrm{exp}}$, and the velocity of the extended material, $v_e$, also enter as parameters. For $M_c$ we use the He core mass from our model of the main bolometric LC peak (Sect.~\ref{sec:mainmodel}). For the explosion energy we use $E=10^{51}$~erg, which gives $v_e$ that is reasonable\footnote{For an envelope in hydrostatic equilibrium the velocity of the extended material should be lower by a factor $1.4\pm0.1$ compared to the photospheric velocity \citep{2010ApJ...725..904N,2014ApJ...788..193N}.} compared to the velocities derived from the evolution of \brad\ and the H$\alpha$ velocities in our earliest spectra (Fig.~\ref{fig:line_vel}). The result of the MCMC simulation is $M_{env}=4.3^{+0.14}_{-0.13}\times10^{-2}$~\msun, $R_{env}=0.73^{+0.03}_{-0.02}\times10^{14}$~cm [$1050^{+40}_{-30}$~\rsun], and 
$t_{\mathrm{exp}}=2458227.1^{+0.1}_{-0.1}$ (JD). Uncertainties represent 68\% confidence intervals. The range of models allowed are visualized in Fig.~\ref{fig:early_lcs}. These models indicate that the contribution from the cooling envelope to the LCs drops rapidly after 6~d past explosion.

We have also used the model by \cite{2016A&amp;A...589A..53N} to model the early bolometric LC (see Fig.~\ref{fig:bol_lc}). This model is fit by eye, and we are able to achieve an excellent match to the early declining LC for, $M_{env}=5.4\times10^{-2}$~\msun, $R_{env}=0.55\times10^{14}$~cm [$\approx790$~\rsun], with an initial kinetic energy $E_k=0.1\times10^{51}$~erg, and initial thermal energy $E_{th}=0.05\times10^{51}$~erg, and $\kappa=0.3$~cm$^2$g$^{-1}$. While these values give the nicest looking fit for the bolometric LC, models that are closer to the result from the MCMC fit based on the \cite{2014ApJ...788..193N} model are still generally within the error bars on our early bolometric datapoints. The expansion velocity of the outer ejecta in the best fitting \cite{2016A&amp;A...589A..53N} model is $\approx20,000$~km~s$^{-1}$, which is consistent with the early BB evolution and the H$\alpha$ velocity in our earliest spectrum (Fig.~\ref{fig:line_vel}). 
Thus, we conclude that both of these models are generally in agreement. The early emission of ZTF18aalrxas can be very well reproduced as the cooling emission from a low mass and very extended envelope. 

Fig.~\ref{fig:bol_lc} shows that the duration of the first LC peak is markedly shorter in ZTF18aalrxas ($\approx6$~d) compared to SN~2013df ($\approx12$~d). A significant difference in cooling phase duration, in the context of these semi-analytical models, can most easily be explained by a difference in the envelope mass around the progenitor at the time of explosion. Using the \cite{2016A&amp;A...589A..53N} model, we find that $M_{env}=11\times10^{-2}$~\msun\ and $R_{env}\approx370$~\rsun\ is required to model the early (unscaled) bolometric emission of SN~2013df\footnote{We use $E_k=0.3\times10^{51}$~erg, and $E_{th}=0.145\times10^{51}$~erg. These values result in an expansion velocity of  $\approx20,000$~km~s$^{-1}$ in the extended shell, roughly consistent with the H$\alpha$ velocity in early spectra of SN~2013df.}. We are using a higher value for $E(B-V)$, but our values are still comparable to the values ($M_{env}=8.0\times10^{-2}$~\msun, $R_{env}\approx160$~\rsun) derived using the same model for SN~2013df by \cite{2016MNRAS.460.1500S}. By similar reasoning it can be argued that the extended envelope mass and radius of SN~1993J should be very similar to that of ZTF18aalrxas; qualitatively the early LCs behave very similarly, with the turn-over from cooling to \Nif\ power happening at a very similar epoch for both SNe. By fitting the unscaled bolometric LC of SN~1993J we derive $M_{env}\approx6\times10^{-2}$~\msun, $R_{env}\approx500$~\rsun.

In conclusion, the hydrogen envelopes of ZTF18aalrxas and SN~1993J appear to be very similar, while the envelope of SN~2013df is slightly more massive. In Sect.~\ref{sec:oi} we arrived at the same qualitative conclusion, based on our nebular spectroscopy. However, the hydrogen masses were somewhat larger. This is likely a calibration issue of the semi-analyitical LC models; we consider the nebular constraints as more indicative of the total H mass at the time of explosion, since these estimates are significantly less model dependent (e.g., they do not depend on assumptions about kinetic or explosion energies).

\subsection{Modeling lightcurves and spectra}
\label{sec:mainmodel}

Since the LC contribution from the cooling of the extended envelope of ZTF18aalrxas becomes insignificant at 6~d past explosion (still almost 20~d before the main bolometric LC peak), the \Nif\ powered part of the bolometric LC can be modeled independently in order to constrain the explosion energy ($E$), the mass of the He core ($M_{\mathrm{He}}$), and the \Nif\ mass ($M_{\mathrm{Ni}}$).

For this purpose we have used the same model (12C) that we found to agree well with our nebular spectrum of ZTF18aalrxas in Sect.~\ref{sec:oi}. We have calculated the evolution of the 12C model between 1 and 200 days\footnote{In this paper our model comparisons start from 6~d past explosion, since this is where \Nif\ starts to dominate over cooling.} with the MC based non-local thermodynamic equilibrium (NLTE) code \texttt{JEKYLL} \citep{2018A&amp;A...620A.156E}. More precisely, this means that we have taken model 12C from \cite{2015A&amp;A...573A..12J}, rescaled it homologously to 1 day and then evolved it with \texttt{JEKYLL}. This \texttt{JEKYLL} version of the 12C model shows a good agreement with SN 2011dh throughout the evolution, and will be discussed in more detail in Ergon et al. (in prep.). Here we compare it to ZTF18aalrxas, and find that it reproduces the main peak in the bolometric LC of ZTF18aalrxas extremely well, after it is scaled by a factor of 2.2. Since the nickel mass of the 12C model is $M_{\mathrm{Ni}}=0.075$~\msun, his means we can estimate the nickel mass for ZTF18aalrxas to $M_{\mathrm{Ni}}=0.17$~\msun. Furthermore, when combined with the \cite{2016A&amp;A...589A..53N} model for the early cooling emission the full bolometric LC can be reproduced (Fig.~\ref{fig:bol_lc}). The rest of the parameters for the 12C model are $M_{\mathrm{He}}=3.1$~\msun\ and $E=0.68 \times 10^{51}$~erg. 
The photospheric velocity of the 12C \texttt{JEKYLL} model is also well matched to the velocities derived from \ion{Fe}{2}~$\lambda$5169 in ZTF18aalrxas (Fig.~\ref{fig:line_vel}).

The $gri$ LCs for the 12C \texttt{JEKYLL} model are compared to those of ZTF18aalrxas in Fig.~\ref{sec:lcs}. The $g$ and $r$ bands are well matched, with in particular the flattening between 50~d and 100~d past explosion observed in the $g$ band for ZTF18aalrxas also seen in the model $g$-band LC. The $i$ band stays somewhat too bright for a few months following maximum light in the model, but becomes consistent with our observations again when the ejecta are in the nebular phase ($\approx150$~d past explosion). In general, the broad band optical emission from the 12C \texttt{JEKYLL} model appears well matched to ZTF18aalrxas.

A set of simulated spectra\footnote{In Fig.~\ref{fig:spec_evol}, the model spectra have been scaled to the nickel mass of ZTF18aalrxas (factor of 2.2). Furthermore, when the epoch of the model spectrum differs from the observed spectrum of ZTF18aalrxas used in the comparison, the model spectrum has been scaled using the model $r$-band LC, so that synthetic photometry on the model spectrum matches the model $r$-band LC at the epoch of the ZTF18aalrxas spectrum.} based on the 12C 
\texttt{JEKYLL} model are shown in Fig.~\ref{fig:spec_evol}. While the observed and modeled spectra are qualitatively similar, with comparable absorption line velocity evolution (Fig.~\ref{fig:line_vel}), we do note some interesting discrepancies. Most importantly, clear He lines appear earlier in the 12C 
\texttt{JEKYLL} model. Our spectrum of ZTF18aalrxas taken at $26$~d past maximum still has very weak He signatures, while the model has already developed clear \ion{He}{1} $\lambda$7065 emission at 18~d past explosion. In later spectra, the \ion{He}{1} lines continue increasing in strength, and at $32$~d past explosion they are clearly much stronger in the 12C 
\texttt{JEKYLL} model compared to what we see in ZTF18aalrxas. The absorption from H$\alpha$ is also stronger in the 12C 
\texttt{JEKYLL} model compared to our observations of ZTF18aalrxas. However, H$\alpha$ emission is stronger in ZTF18aalrxas. 

These observational facts can likely be explained if the hydrogen envelope of ZTF18aalrxas is more massive compared to that of the 12C model, such that the He emission is blocked for a longer time early on\footnote{Some adjustment of the \Nif\ mixing throughout the ejecta could also result in a more delayed appearance of the He lines.}. The H$\alpha$ velocities in the 12C 
\texttt{JEKYLL} model are also somewhat faster, indicative of a lower H mass (for the same explosion energy). A higher H mass, especially in the extended envelope of the progenitor, is also supported by the early cooling emission that is much stronger in ZTF18aalrxas compared to SN~2011dh. 

As the ejecta become optically thin, and the signatures from H and He decrease, the 12C \texttt{JEKYLL} model becomes very well matched to ZTF18aalrxas, except for possibly some excess H$\alpha$ (see the spectrum taken at 177~d past explosion, and the discussion in  Sect.~\ref{sec:ha}). This should not be surprising since the 
same ejecta model is used as in the original 12C model by \cite{2015A&amp;A...573A..12J} that we investigated in Sect.~\ref{sec:oi}. The O mass in the 12C model is 0.3~\msun, and the ZAMS mass is 12~\msun.

In conclusion, the 12C model, while it could use some fine tuning of the structure and mass of the H envelope, can be used to robustly constrain the \Nif\ mass and He core mass of ZTF18aalrxas.

\section{Discussion and Conclusions}\label{sec:conclusions}
We have performed a thorough analysis of the optical emission from ZTF18aalrxas, finding that the progenitor of this SN had, in the context of SE SNe (e.g., \citealp{2018A&amp;A...609A.136T}), a low mass ($M_{ZAMS}=12-13$~\msun) and was surrounded by a very extended envelope ($R_{env}=790-1050$~\rsun), with a mass ($M_{env}=0.04-0.15$~\msun) at the time of explosion.

Our ZAMS mass results are based on the 12C model, originally presented by \citet{2015A&amp;A...573A..12J}, and here extended in time with a new MC NLTE code (\texttt{JEKYLL}; \citealp{2018A&amp;A...620A.156E}). The 12C \texttt{JEKYLL} model is able to reproduce the main bolometric LC of ZTF18aalrxas very well, while at the same time also reproducing the [\ion{O}{1}] $\lambda\lambda$6300, 6364 emission in our nebular spectrum of the SN. These two constraints are usually obtained from separate modeling codes. One that simulates the bolometric LC, another that simulates the nebular phase spectra, and these simulations in general do not use the same progenitor model. Our new 12C \texttt{JEKYLL} model offers strong evidence for a very low ZAMS-mass progenitor for ZTF18aalrxas in the $12-13$~\msun\ range, consistent with the ZAMS mass predicted for SE SNe by binary evolution modeling (e.g., \citealp{2017ApJ...840...10Y}), and not consistent with SE SNe produced through wind-driven mass loss from massive stars (e.g., \citealp{Groh:2013ab}). Thus, it is very likely that the progenitor of ZTF18aalrxas was part of a binary system. 

Furthermore, our nebular spectrum of ZTF18alrxas shows an excess around the H$\alpha$ line, which can be explained through interaction between the SN ejecta and a dense CSM (Sect.~\ref{sec:ha}). Incidentally, all of the well studied SNe IIb in the literature that show very strong cooling signatures in their early LCs (due to their large radii at the time of explosion), also show signs of CSM interaction in their nebular spectra (SN~1993J, SN~2013df). While not as well studied, this is also likely the case for SN~2011fu (\citealp{2013MNRAS.431..308K,2015MNRAS.454...95M}, although the interpretation of this excess is different in \citealp{2015MNRAS.454...95M}). In contrast, no dense CSM (no significant excess around the H$\alpha$ region in nebular spectra) is generally seen in SNe IIb that are more compact at the time of explosion and lack strong cooling signatures (e.g., SN~2011dh, SN~2008ax; see \citealp{2015A&amp;A...573A..12J}, and PTF12os; see \citealp{2016A&amp;A...593A..68F}). Thus, as already suggested by  \cite{2010ApJ...711L..40C}, it appears that there indeed are two distinct classes of SNe IIb; one compact class, with less massive H envelopes without a dense CSM, and one extended class with more massive H envelopes and a dense CSM (see also \citealp{2015ApJ...803...40B}). Here we have added another member, ZTF18aalrxas, to the extended envelope SN IIb family. 

In binary model systems for SE SNe, the mass-loss rate (and the potential for the presence of a dense CSM) is driven by the binary interactions (e.g., \citealp{2014ARA&amp;A..52..487S}). Based on this, \cite{2015ApJ...807...35M} proposed that the extended and compact classes of SNe IIb represent two different binary evolution paths. Extended SNe IIb (such as SN~1993J) would be exploding when strong binary interaction and subsequent mass loss is ongoing, while compact SNe IIb (such as SN~2011dh) should have a significant delay between the mass loss episode where most of the H envelope is lost and the explosion.

More recently, \cite{2017ApJ...840...10Y} have constructed a large grid of binary models for SNe IIb, and the results from this work at least partly disagree with the picture suggested by \cite{2015ApJ...807...35M}. Compact SNe IIb are in \cite{2017ApJ...840...10Y} produced from systems in a tight orbit undergoing early Case B mass transfer, which means there is indeed a significant delay between the strong interaction and mass loss phase and the explosion. However, \cite{2017ApJ...840...10Y} also show that extended SNe IIb could be produced through wider orbits which result in late Case B systems, where there is also a significant delay until the explosion. Some of these models for extended SNe IIb also seem to produce strong enough winds to be roughly consistent with the presence of CSM interaction (the high mass-loss rate is a function of the radius and final mass at the time of explosion). However, we note that in the current grid, stars with ZAMS masses in the range suitable for ZTF18aalrxas ($M_{ZAMS}=12-13$~\msun), will not have strong enough mass loss at the time of explosion to explain the CSM signature; a model with $M_{ZAMS}=13$~\msun, at solar metallicity, and a final H mass of roughly 0.15~\msun\ (consistent with our nebular spectra), will result in a progenitor with $R\approx640$~\Rsun\ and $\dot{M}\sim4\times10^{-6}$~\msun~yr$^{-1}$. The radius of this model is indeed close to what we derive from the early cooling emission of ZTF18aalrxas, but the wind mass loss is around an order of magnitude too low when compared to the mass-loss rates derived for SN~2013df and SN~1993J. Suitable mass-loss rates are only possible for progenitors with $M_{ZAMS}>16$~\msun\ in the grid by \cite{2017ApJ...840...10Y}, which is not consistent with our LCs and nebular spectrum of ZTF18aalrxas. It is possible that a large enough parameter space has not yet been explored; and ongoing mass transfer might really be needed to have high enough mass loss in the lower mass progenitor systems. It is possible to have binary systems that produce SN IIb explosions after multiple episodes of mass transfer during both He and later burning stages (see e.g., \cite{2013ApJ...762...74B}, although this model produces a compact SN IIb). 

We propose that nebular spectroscopy can be used to test this issue; if a Case B scenario like in \cite{2017ApJ...840...10Y} dominates the production of extended SNe IIb, there should be a relation between the ZAMS mass, the CSM density, and the final envelope mass and radius; and a continuum in the strength of the CSM interaction should be seen. If an episode of strong binary interaction must be ongoing at the time of explosion, a high CSM density should be possible for both low and high progenitor $ZAMS$ masses, and the CSM interaction should either be non-existent or strong; there should not be a continuum between compact and extended SNe IIb in terms of the CSM driven late-time H$\alpha$ emission.

The ZAMS mass we find for ZTF18aalrxas is similar to that found for SN~2011dh, which is on the lower end for what has been observed in SE SNe, based on their bolometric LCs (e.g., \citealp{2018A&amp;A...609A.136T}), and also their late-time oxygen emission \citep{2015A&amp;A...573A..12J}. The duration of the first LC peak is somewhat shorter in ZTF18aalrxas ($\approx6$~d) compared to SN~2013df ($\approx12$~d), and we interpret this difference as a lower extended envelope mass for the progenitor of ZTF18aalrxas at the time of explosion compared to SN~2013df (Sect.~\ref{sec:early_lcs}). We also found that SN~1993J appears to have an extended envelope nearly identical in mass and extent to that of ZTF18aalrxas. Furthermore, a similar conclusion was also reached based on the late-time H$\alpha$ emission of these SNe (Sect.~\ref{sec:ha}). The progenitor of SN~1993J is likely the most massive among the progenitors to these extended SNe IIb (based on late-time O emission; \citealp{2015A&amp;A...573A..12J}), but it did not have the strongest cooling signature. SN~2011fu is well matched to a $13$~\msun\ ZAMS mass model \citep{2015MNRAS.454...95M}, and had a cooling signature even stronger than SN~2013df. Thus, while the sample is small, it appears that there is currently no relation between ZAMS mass and the strength or duration of the cooling emission in extended SNe IIb; the ZAMS masses instead appear to be similar in extended SNe IIb and compact SNe IIb. This could be argued to be in support of the need for an ongoing mass loss episode at the time of explosion. However, the current sample of objects is far too small to make any real claims; a systematic study of the nebular spectra of a larger sample of SNe IIb is needed.

In conclusion, it appears clear that ZTF18aalrxas, SN~1993J, SN~2013df are all very similar; they all show strong long-lasting cooling signatures in their early LCs, and over time they develop very similar CSM powered flat-topped H$\alpha$ profiles in their spectra - indicative of the presence of dense CSM. They likely originate from binary systems with very similar configurations, that result in strong mass loss during the final centuries before the explosions (either as a result of strong winds, or ongoing interaction). We note that as CSM can already be identified clearly at 177~d, it is also likely affecting earlier observations; the result of a dense CSM is generally shallower P-Cygni profiles (see e.g., \citealp{2017A&amp;A...605A..83D}), which is exactly what is seen in SN~1993J, SN~2013df, and ZTF18aalrxas, when compared to SN IIb models without any CSM (Fig.~\ref{fig:spec_evol}). This also implies that the CSM could be affecting the bolometric luminosity of these SNe, which would lead to overestimated \Nif\ masses when they are modeled without taking this into account. This possibility deserves significant future consideration.

\acknowledgments
Based on observations obtained with the Samuel Oschin Telescope 48-inch and the 60-inch Telescope at the Palomar Observatory as part of the Zwicky Transient Facility project. ZTF is supported by the National Science Foundation under Grant No. AST-1440341 and a collaboration including Caltech, IPAC, the Weizmann Institute for Science, the Oskar Klein Center at Stockholm University, the University of Maryland, the University of Washington, Deutsches Elektronen- Synchrotron and Humboldt University, Los Alamos National Laboratories, the TANGO Consortium of Taiwan, the University of Wisconsin at Milwaukee, and Lawrence Berkeley National Laboratories. Operations are conducted by COO, IPAC, and UW. 

This work was supported by the GROWTH project funded by the National Science Foundation under PIRE Grant No 1545949. The Oskar Klein Centre is funded by the Swedish Research Council.

Part of this research was carried out at the Jet Propulsion Laboratory, California Institute of Technology, under a contract with the National Aeronautics and Space Administration.

Partially based on observations made with the Nordic Optical Telescope, operated by the Nordic Optical Telescope Scientific Association at the Observatorio del Roque de los Muchachos, La Palma, Spain, of the Instituto de Astrofisica de Canarias. Some of the data presented here were obtained with ALFOSC, which is provided by the Instituto de Astrofisica de Andalucia (IAA) under a joint agreement with the University of Copenhagen and NOTSA.

Some of the data presented herein were obtained at the W. M. Keck
Observatory, which is operated as a scientific partnership among the
California Institute of Technology, the University of California, and
NASA; the observatory was made possible by the generous financial
support of the W. M. Keck Foundation. 

This paper is partly based on observations made with the Italian Telescopio Nazionale Galileo (TNG) operated on the island of La Palma by the Fundaci{\'o}n Galileo Galilei of the INAF (Istituto Nazionale di Astrofisica) at the Spanish Observatorio del Roque de los Muchachos of the Instituto de Astrofisica de Canarias. Partially based on observations obtained with the Apache Point Observatory 3.5-meter telescope, which is owned and operated by the Astrophysical Research Consortium.

The Liverpool Telescope is operated on the island of La Palma by Liverpool John Moores University in the Spanish Observatorio del Roque de los Muchachos of the Instituto de Astrofisica de Canarias with financial support from the UK Science and Technology Facilities Council.

The SED Machine is based upon work supported by the National Science Foundation under Grant No. 1106171. 

A.G.-Y. is supported by the EU/FP7 via ERC grant 307260, ``The Quantum Universe'' I-Core program by the Israeli Committee for planning and budgeting, by ISF,
GIF, and Minerva grants, and by the Kimmel award.

This work is part of the research programme VENI, with project number 016.192.277, which is (partly) financed by the Netherlands Organisation for Scientific Research (NWO).

\bibliography{18aalrxas}

\end{document}